# FUTURE ROBOTICS DATABASE MANAGEMENT SYSTEM ALONG WITH CLOUD TPS

Vijaykumar S and Saravanakumar S G

Thiagarajar School of management
Thirupparankundram
Madurai 625 005, Tamil Nadu, INDIA
Email: indianid@gmail.com.
Team id: Project6thsense@googlegroups.com
http://www.indianid.6thsense.us
Sastra University
Kumbakonam-612001
Tamil Nadu, INDIA
Email: saravanakumarsg@gmail.com

## ABSTRACT

*This paper deals with memory management issues of robotics. In our proposal we break one of the major issues in creating humanoid. . Database issue is the complicated thing in robotics schema design here in our proposal we suggest new concept called NOSQL database for the effective data retrieval, so that the humanoid robots will get the massive thinking ability in searching each items using chained instructions. For query transactions in robotics we need an effective consistency transactions so by using latest technology called CloudTPS which guarantees full ACID properties so that the robot can make their queries using multi-item transactions through this we obtain data consistency in data retrievals. In addition we included map reduce concepts it can splits the job to the respective workers so that it can process the data in a parallel way.*

## 1. INTRODUCTION

In the modern age robotics, we are facing very big problem to manage and retrieve information. When the embedded system we can only able to perform some task only like playing football robots, manufacturing robot, etc.in embedded system we are using multiple sensor to do various task and from sensor we give specific task for robots. But when the matter comes to making a robot like human beings humanoids we have to consider n number of things like intelligence up to instructions. Consider an middle finger have tree joints for that we want to give tree part of instruction and each part has ten states fuzzy sets (0to1) for the movement and these have property like movement fast this way for each items we have to manipulate billions of instruction instead of that we go with an sensor on that case also we have an minimize the instruction regarding to that we want alternate solution to done an massive process and with massive intelligence. How it is possible? Because for that we need huge amount of memory, processing capacity, huge instruction, upgradable capability, storage memory and a final important thing is data management. These are the key blocking things in the humanoid making process. But today we have an ability to achieve this by using a techniques and technology like NOSQL, fuzzy logic, Map Reduce, etc. Here we take a challenge to make a humanoid with amazing intelligence with massive processing technique. From this method we can able to store and retrieve yota byte (1024) of information. Using this NOSQL we can able to retrieve that information in a quicker and efficient manner.

## 1.1. Database

A database is any collection of related data. And the restrictive of a database is a persistent, logically coherent collection of inherently meaningful data, relevant to some aspects of the real world [26].

## 1.2 Database Management System

A database management system (DBMS) is a collection of programs that enables users to create and maintain a database. According to the ANSI/SPARC DBMS Report (1977) [26].

## 1.3 Relational database

A database that treats all of its data as a collection of relations and the characteristics of relations are [26]. A kind of set, a subset of a Cartesian product and an unordered set of ordered tuples

## 1.4 Problem with RDBS

The important problem with a RDBMS is difficult to scale bulk amount of data. they have facing 3 TB for "Green Badges", on that way Facebook handles 100 TB for inbox search and EBay handles 2PB and twitter handles 2PB every day for user images so the relational base are difficult to handle this much amount of data due to rigid schema design is the cause for this failure and we know server crash also happen due to data management sometimes the geo informatics service server also crash because of DBMS failure think it is the small amount of information when we compare with an humanoid knowledge information because maps are part of an humanoid information because the situation may occur to store more than 500TB to store human face images but in RDBMS how do we able update human face changes it is also an major restriction .If a relational database running a query which has multiple views with multiple inner and outer joins, grouping, summing and averaging against tables with 90 million rows then result will not appropriate for these type of queries because the processing time is very high and it difficult scale over such huge tables.

RDBMS use timestamp for issue tracking purpose they will check the time stamp for the delivered messages but in NOSQL we use timestamp for current data retrievals.

## 2. Definition for NOSQL

Next Generation Databases address some of the following being non-relational, distributed, open-source and horizontal scalable more nodes can be added. The original intention has been modern web-scale databases. NOSQL was first developed in the late 1990's by CarloStrozzi. The movement began early 2009 after it's growing rapidly.

## 2.1 Why NOSQL?

"NoSQL" is a general name for the collection of databases that do not use SQL (Structured Query Language) or a relational data model. It did not say that the relational database are bad choice but is says we do not think that is the only choice for data storage. [37

## 2.2 Translation Table

 This translation table explains you to know the NOSQL properties by its equivalent older meaning

Table 1: Keyword Translation

| OLD NAME | NEW NAME |
|---|---|
| Hash file | Key-Value Store |
| Hierarchical file (HSAM, HDAM) | BigTable |
| Parent node | Column family |
| Local autonomy | Partition tolerant |
| Horizontal partition | Sharding |
| non-ACID (atomic, consistent, isolated, durable) | BASE (basically available, soft state, eventually consistent) |

## 2.3 Characteristics

NOSQL normally doesn't have an ACID property like (atomicity, consistency, isolation, durability), no join operation, special of the NOSQL is schema-free, replication support, easy API, eventually consistency, and more. So the misleading term "NOSQL" (the community now translates it mostly with "Not Only SQL"). And it is structured storage and usually has a collection of tables with structured data (most probably like a hash table or a dictionary) then no need to map object-oriented designs into a relational model.

**Examples**    Google's BigTable, Amazon's Dynamo.

Cassandra (used in Facebook's inbox search) and

HBase (Apache) are open source

## 2.4 CAP Theorem and NOSQL

**Fig.1.** CAP Theorem Satisfaction

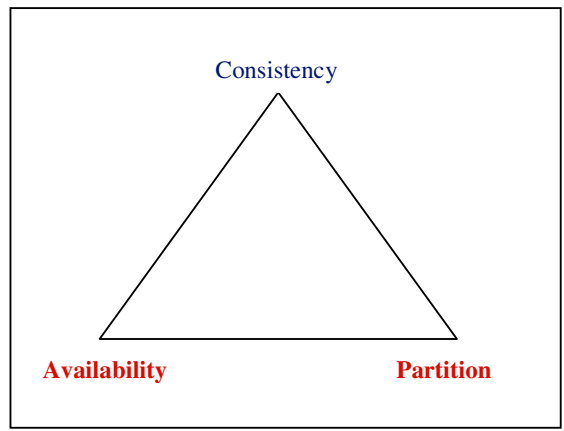

**[28]***CAP (FOR NOSQL DATABASES)( FOR EASY SCALABILITY)*
- CONSISTENCY: All database clients see the same data, even with concurrent updates.
- AVAILABILITY: All database clients are able to access same version of the data and easy scalability
- PARTITION TOLERANCE: The database can be split over multiple servers.

### 2.5 Core NOSQL Systems

NOSQLS Systems where many in types but these where the core types of NOSQL Systems
1. Store / Column Families
2. Document Store
3. Key Value / Tuple Store
4. Eventually Consistent Key Value Store
5. Graph Databases

### 2.6 Data And Query Model

There is a lot of variety in the data models and query APIs in NOSQL databases.

Table 1: Data and Query Model Table

|  | Data Model | Query API |
|---|---|---|
| Cassandra | Column family | Thrift |
| CouchDB | Document | Map/reduce views |
| HBase | Column family | Thrift,REST |
| MongoDB | Document | Cursor |
| Neo4J | Graph | Graph |
| Redis | Collection | Collection |
| Riak | Document | Nested hashes |
| Scalaris | Key/Value | Get/put |
| Tokyo Cabinet | Key/Value | Get/put |
| Voldemort | Key/Value | Get/put |

### 2.7 Persistence Design

Table 2: Data Storage Design Table

|  | **Persistence Design** |
|---|---|
| Cassandra | Memtable/SSTable |
| CouchDB | Append-only B-tree |
| HBase | Memtable/SSTable on HDFS |
| MongoDB | B-tree |
| Neo4J | On-disk linked lists |
| Redis | In-memory with background snapshots |
| Scalaris | In-memory only |
| Tokyo Cabinet | Hash or B-tree |
| Voldemort | Pluggable(primarily BDB MySQL) |

This are many number of persistence design avail today but above I give some famous model. It gives you a various choice to implement NOSQL depend on your need here below I take one GRAPH model Neo4J to give an view about that from that we can able to analyses the scenario and implement on it.

## 2.8 Tablet Location

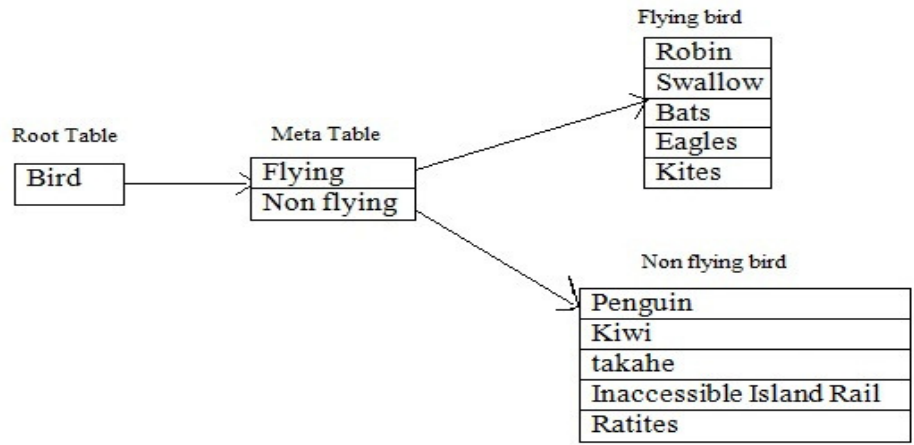

The above figure clearly shows how the information is splited in different tables, in the root table has a reference to its respective meta tables then meta tables has meta information about that query then it refers to the particular user table where the full brief description is given. The humanoid will cache the table locations if they find location is incorrect they will search using chubby file and it traverse though root table. NoSQL has log files it stores information's about the previous transactions so that when a system is searching for a new query it can easily refer previous log files to get some idea, So that the query is processed without looking for the each tables.

## 3  Graph Model

Graph database it stores the value of nodes, edges and properties. There are some general graph database are available that stores any graph and some special kinds of graph database are also available like triple store and network database. In network database it uses edges and nodes to represent and store the data. Graph database is faster when compare to the relational database it map more directly to the structure of object-oriented applications And they successfully implemented in.

- Social networking
- Represent the real world

Is the one of the best NOSQL type to make mind mapping from that we can able mapping the brain and forming a fuzzy based intelligence.

**EXAMPLE**

Node firstNode = graphDb.createNode();
Node secondNode = graphDb.createNode();
Relationship relationship = firstNode.createRelationshipTo(secondNode,
MyRelationshipTypes.KNOWS );
eg. Neo4j
firstNode.setProperty( "message", "Arun, " );
secondNode.setProperty( "message", "Raju" );
relationship.setProperty( "message", " son" );
The graph will look like this:

(firstNode )---KNOWS--->(secondNode)
*Printing information from the graph:*
System.out.print( firstNode.getProperty( "message txt" ) );
System.out.print( relationship.getProperty( "message txt" ) );
System.out.print( secondNode.getProperty( "messagetxt" ) );

eg. Neo4j
Printing will result in:
Arun son Raju

### 3.1   How far the Graph Model Helps To Effective Data Retrieval in Robots

In Triple store database it can store triple meaningful data but in the humanoid we need to process with large grouping of data in such cases graph database provides more convenient way of approach in storing the data's in the nodes and its properties describes how it can be search using the queries and the edges shows the relationship among the nodes so that it can form any group based on the query data. As we discussed earlier the humanoid queries are chained so that graph approach provides quicker access without join operation, the humanoid can apply any type of searching mechanism in finding the data's. the results of graph database are different when compare to the other database here we can provide Provenance query answerability this approach is useful when there is a situation in need of making comparative analysis this method can be applied, in this method we can suggest the answer of second question by analyzing the first result. This concept is helpful in decision making in humanoid.

### 3.2 Decision making using the Graph Model:

 Graph model not only used for data retrievals it also used for the decision making in humanoids. At first it will search for the obstacles in that path then it will map those items in a graph model they will assign node weight according the distance and obstacles weight then they will find best shortest path algorithm to reach designation point in such way the humanoid makes the decisions by using the graph model. We can use algorithms such as A*, Dijkstra's algorithm.

## 4   OUR PROPOSAL SYSTEM NOSQL ON HUMANOID BRAIN
**Fig. 2.** IMPLEMENTATION OF NOSQLON HUMANOID ARTIFICIAL BRAI

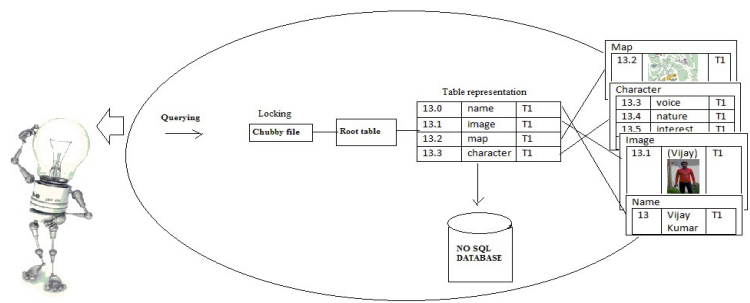

## 4.1 Chained queries:

Cassandra
super columns:
```
{
   "column family 1": {
     "super key 1": {
       "key 1": {
          "property 1": "value",   Time stamp:"Value"
          "property 2": "value",   Time stamp :"value"
       },
       "key 2": {
          "property 1": "value",   Time stamp:"Value"
          "property 4": "value",   Time stamp:"Value"
          "property 5": "value",   Time stamp:"Value"
       }, ...
     }, ...
     "super key 2": {
       "key 1": {
          "property 4": "value",    Time stamp:"Value"
          "property 5": "value"     Time stamp:"Value"
       },
       "key 2": {
          "property 1": "value",   Time stamp:"Value"
          "property 6": "value",   Time stamp:"Value"
          "property 7": "value",   Time stamp :"Value"
       },
     },
   },
}
```
In the above example clearly shows how the chained queries are processed using two key values in NOSQL we can frame queries using two search key elements so that one operation can have two set of results if it is matches with the second key then it goes with that branch. The property value can be anything related to that column families stored in the table it can be link to another webpage or image.

## 4.2 Time stamps:

  In this model we are used 128 bit integer timestamp.
Timestamp is the important thing in the data retrievals in humanoid so we use two columned time stamp mechanism.
It has two columns
1. Index timestamps is the super column (real time in microseconds)
2. Index for the timestamp index (T1, T2,"com.cnn.www");

**Examples:**
 Timestamp super 1
```
{
        T1
          {   Value "com.cnn.www"   }
        T2
          {     Value "com.cnn.www"   }
}
```

### 4.3 Cloud TPS:

It is the middleware layer for NOSQL which guarantees full ACID properties for multi-item transactions. The architecture describes that each LTM (Local Transaction Managers) has the replica of its query to the other LTM's, to support ACID guarantees it works on 2-Phase Commit protocol (2PC) so that if all the LTM's commits the transactions successfully then its coordinator will complete the transactions else it will aborted the transactions. For the simple consistent read operations the query read directly in the cloud data service without referring the LTMs. [36]

### 4.4 Cloud TPS on robotics:

Key differences between cloud data services

|  | **SimpleDB** | **Bigtable** |
| --- | --- | --- |
| **Data Item** | Multi-value attribute | Multi-version with timestamp |
| **Schema** | No schema | Column-families |
| **Operation** | Range queries on arbitrary attributes of a table | Single-table scan with various filtering conditions |
| **Consistency** | Eventual consistency | Single-row transaction |

### 4.5 Humanoid Functionality

Robot electronic system it can't recognize human speech and image. it can repose only to the binary number. Generally the binary digits are eight bits in length. In robot instructions are spited into many pieces and stored in many places because the instruction are in a form of chains, if one instruction starts it continued by another instruction, to explain briefly the robot parts are divided into pieces, for example take a hand it has following parts modified spiral joint, revolute joints, spherical joint, phalanges, knuckles etc. if we want to take an object or a particle we need to move all these parts to perform the given task and we need an instructions need to be frame it. If we pass multiple pieces of Instructions to move a hand itself we need a multiprocessor system to process all those instructions, assume that the robot has various parts that made to work to perform a task. The situation is seemed to be more complex, to resolve this conflict we going to provide a solution for this, instead of passing group of instructions, the instructions are passed based on the task it will first call one instruction and it address another instruction then it continues until all instructions are passed if we want to follow this mechanism we need an effective database to handle with this much of instruction.

**Fig.3.** NOSQL DB Model for Robotic instruction

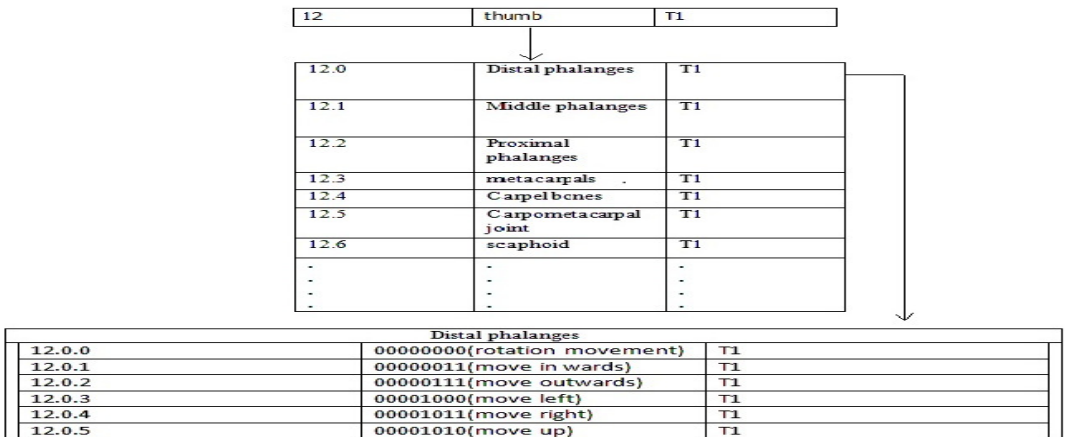

### 4.6 More memory using NOSQL

Nowadays NOSQL is the popular non-relational database it can handle with terabytes of data, it has an time stamp mechanism so that it queries current data without need of any special query to retrieve the latest information, it is very helpful in the robots because the robots will scan and updates the data's regularly so that the time stamp mechanism helps to reduce the processing time. In NOSQL database it has a column family's concept the Column keys are grouped into sets called column families, which form the basic unit of access control. All data stored in a column family is usually of the same type. By using this concept we spilt the instructions based on the task and the portion need to be moved, so that we can easily organize the data's (instructions) in column families, so that with the help of one object we can easily refer the entire object based on the task.

**Fig.4.** NOSQL BD Model for Robotic instruction
Information: Grouping of instruction for thumb (back node)

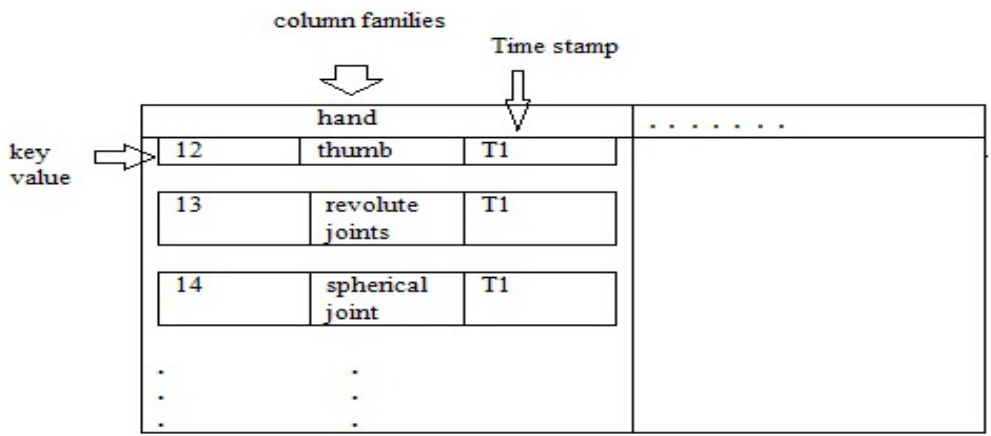

## 5   MAP REDUCES IN ROBOTICS

Map reduce is the framework for processing the large problems, it is need in robots because robots can scan large image and it will try to stored it in database at that time the Database finds difficulty to break the image into pieces and store the respected set. If suppose the robot made a query to Match the current scanning image with the database at that time the database find difficulty to combine the data that are stored, so that it need to focus more on queries to remove all this drawbacks map reduce was introduced it helps to divide the problems into many pieces and it given to the several worker node.

### 5.1   Master Node

The master node takes the input and it assigns the work to the clients.

**Fig.5.** MAP REDUCE PROPOSAL SYSTEM MODEL ON ROBOTICS

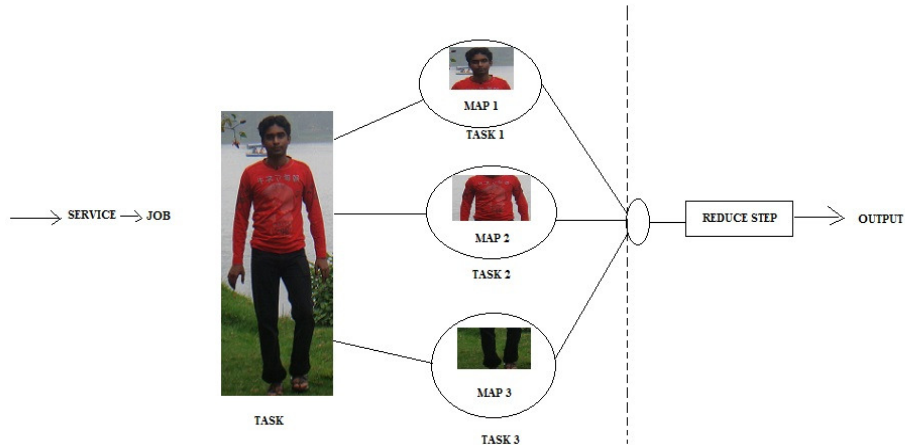

## 5.2  Reduce Step

In the reduce step the result are combined and given to the master node from that figure 5 you can easily get idea about that.

## 5.3  Importance

Above will explain you about the data management and retrieve capacity of NOSQL but this Map Reduce give a massive performance in terms of input and output process from NOSQL.

# 6  Advantages of Our NOSQL Implementation on Robotics

NOSQL is schema-free database's so it is easy to implement and maintain, it can scale up and down, these database's are replicated to avoid fault-tolerant and can be partitioned if it scales large, the data are easily distributed to the database's, it can process large amount data within a short period of time, it supports specific problem/situation that are no need to think in terms of relations but in terms given in a situation (e.g. documents, nodes,...) in most cases it is freely available because in most of the products are open-source.

First we don't go with huge data server so it gives consume cost and it also have ability to consume power, from this proposal you can able to achieve superfast because we can split the input and Output easily using this map reduce concept from that we easily go with cloud computing.

Then the concept time stamp we easily update our robots without any redundancy of data so we easily avoid garbage collection and give automation to unwanted information deletion and the important thing is the updating process happens immediately when you do this. E.g. like updating the cricket score after updating it immediately publishes that info and replace the old score. It has ability to handle billions of objects so we improve the vision intelligence and hearing intelligence so it is the major step to produce humanoid with very high sensitive intelligence. E.g. the database used for Amazon S3, which as of March 2010 was hosting 102 billion objects.

## 7   Data Retrievals and Meaningful Schemas Issues in Database

In the relational database design the table schemas are framed by applying the 3 normal forms so that it did not have data redundancy issues while retrieving the data sets and the each table contains meaningful schemas. But in NOSQL it did not focus on what kind of data it stores and how the design should be it entirely focus on how it can be stored effectively. In NOSQL the data sets are divided into multiple pieces of information and each carries its own time stamps based on the arrivals. By using the map reduce it stores the data in a convenient way then the question arises how the data can be retrieved? By using the column families it gives what kind of data it stores based on the some relations and through the time stamp mechanism we obtain the current data. NOSQL supports garbage collections removal so that the old time stamped messages can be removed easily. We already said that humanoid will use its instructions based on fuzzy sets so that instructions can be linked in a chained manner. The chained instructions is given to the map it splits its job to the respective workers in reduce step so that we can process chained queries in such a way that the humanoid can think.

## 8   CONCLUSIONS

In future our database concept can be major part in making robotics it brings an effective way in making the robots to think. This database concept can be applied in distributed environment so that the data can be maintained in separate place and the updates can be done through networking. From NOSQL we can easily achieve the new era of humanoids.

### ACKNOWLEDGMENTS

The authors are grateful to Dr. S. Kannan Advisor of project6thsense and to all team members of project6thsense for the making of our entire research projects and for your valuable feedbacks.

## REFERENCES

[1]  ABADI, D. J., MADDEN, S. R., AND FERREIRA,M. C. Integrating compression and execution in columnoriented database systems. *Proc. of SIGMOD* (2006).
[2]  AILAMAKI, A., DEWITT, D. J., HILL, M. D., AND SKOUNAKIS,M. Weaving relations for cache performance. In *The VLDB Journal* (2001), pp. 169.180.
[3]  BANGA, G., DRUSCHEL, P., AND MOGUL, J. C. Resource containers: A new facility for resource management in server systems. In *Proc. of the 3rd OSDI* (Feb. 1999), pp. 45.58.
[4]  BARU, C. K., FECTEAU, G., GOYAL, A., HSIAO,H., JHINGRAN, A., PADMANABHAN, S., COPELAND,G. P., AND WILSON, W. G. DB2 parallel edition. *IBM Systems Journal 34*, 2 (1995), 292.322.
[5]  BAVIER, A., BOWMAN, M., CHUN, B., CULLER, D.,KARLIN, S., PETERSON, L., ROSCOE, T., SPALINK, T., AND WAWRZONIAK, M. Operating system support for planetary-scale network services. In *Proc. of the 1st NSDI* (Mar. 2004), pp. 253.266.
[6]  BENTLEY, J. L., AND MCILROY, M. D. Data compression using long common strings. In *Data CompressionConference* (1999), pp. 287.295.
[7]  BLOOM, B. H. Space/time trade-offs in hash coding with allowable errors. *CACM 13*, 7 (1970), 422.426.
[8]  BURROWS, M. The Chubby lock service for loosely coupled distributed systems. In *Proc. of the 7th OSDI* (Nov. 2006).
[9]  CHANDRA, T., GRIESEMER, R., AND REDSTONE, J. Paxos made live . An engineering perspective. In *Proc. of PODC* (2007).
[10] COMER, D. Ubiquitous B-tree. *Computing Surveys 11*, 2 (June 1979), 121.137.
[11] COPELAND, G. P., ALEXANDER, W., BOUGHTER, E. E., AND KELLER, T. W. Data placement in Bubba. In *Proc. of SIGMOD* (1988), pp. 99.108.
[12] DEAN, J., AND GHEMAWAT, S. MapReduce: Simpli_ed data processing on large clusters. In *Proc. of the 6th OSDI* (Dec. 2004), pp. 137.150.
[13] DEWITT, D., KATZ, R., OLKEN, F., SHAPIRO, L., STONEBRAKER, M., AND WOOD, D. Implementation techniques for main memory database systems. In *Proc. of SIGMOD* (June 1984), pp. 1.8.
[14] DEWITT, D. J., AND GRAY, J. Parallel database systems: The future of high performance database systems.*CACM 35*, 6 (June 1992), 85.98.
[15] FRENCH, C. D. One size _ts all database architectures do not work for DSS. In *Proc. of SIGMOD* (May 1995), pp. 449.450.

[16] GAWLICK, D., AND KINKADE, D. Varieties of concurrency control in IMS/VS fast path. *Database Engineering Bulletin 8*, 2 (1985), 3.10.
[17] GHEMAWAT, S., GOBIOFF, H., AND LEUNG, S.-T. The Google _le system. In *Proc. of the 19th ACM SOSP* (Dec. 2003), pp. 29.43.
[18] GRAY, J. Notes on database operating systems. In *OperatingSystems . An Advanced Course*, vol. 60 of *Lecture Notes in Computer Science*. Springer-Verlag, 1978.
[19] GREER, R. Daytona and the fourth-generation language Cymbal. In *Proc. of SIGMOD* (1999), pp. 525.526.
[20] HAGMANN, R. Reimplementing the Cedar _le system using logging and group commit. In *Proc. of the 11th SOSP* (Dec. 1987), pp. 155.162.
[21] HARTMAN, J. H., AND OUSTERHOUT, J. K. The Zebra striped network _le system. In *Proc. of the 14th SOSP* (Asheville, NC, 1993), pp. 29.43.
[22] KX.COM. kx.com/products/database.php. Product page.
[23] LAMPORT, L. The part-time parliament. *ACM TOCS 16*,2 (1998), 133.169.
[24] MACCORMICK, J., MURPHY, N., NAJORK, M., THEKKATH, C. A., AND ZHOU, L. Boxwood: Abstractions as the foundation for storage infrastructure. In *Proc.of the 6th OSDI* (Dec. 2004), pp. 105.120.
[25] MCCARTHY, J. Recursive functions of symbolic expressions and their computation by machine. *CACM 3*, 4 (Apr.1960), 184.195.
[26] Database Fundamentals. Robert J. Robbins, Johns Hopkins University, rrobbins@gdb.org
[27] NOSQL-databases.org
[28] http://books.couchdb.org/relax/intro/eventual-consistency
[29] http://www.rackspacecloud.com/blog/2009/11/09/NOSQL-ecosystem/.Jonathan Ellis.
[30] http://blog.evanweaver.com/articles/2009/07/06/up-and-running-with-cassandra/
[31] http://horicky.blogspot.com/2009/11/NOSQL-patterns.html,
[32] http://www.slideshare.net/Eweaver/cassandra-presentation-at-NOSQL
[33] Bigtable:A Distributed storage system for structured data ,google,Inc,OSDI 2006.Map Reduce: Simplified Data processing on large cluster, Google, Inc.
[34] Vijay Kumar S, Saravanakumar S.G., Revealing of NOSQL Secrets. CiiT Journal.vol2,no10(Oct.2010),310314.DOI=http://www.ciitresearch.org/dmkeoctober2010.html
[35] Vijay Kumar S, Saravanakumar S.G., Emerging Trends in Robotics and Communication Technologies (INTERACT),2010. (Dec.2010). DOI=http://ieeexplore.ieee.org/xpl/freeabs_all.jsp?arnumber=5706225
[36] CloudTPS: Scalable Transactions for Web Applications in the Cloud Zhou Wei, Guillaume Pierre, chi-hung chi ieee transactions on services computing, special issue on cloud computing, 2011
[37] http://io.typepad.com/glossary.html

**Biography**

**Vijay Kumar** born in 1989 (16-02-1989) in Kumbakonam, Tamil Nadu, India. He has completed B.Sc. Degree in Computer Science from PRIST University, Tamil Nadu with First Class.  He has completed his P.G. Diploma in Information and Communication Laws from Madurai Kamaraj University, Tamil Nadu.  Currently he is pursuing his MCA Degree in Thiagarajar School of Management, Madurai, TamilNadu in India. He has to his credit more than 18 research publications in various reputed national and International journals. He has communicated over 22 research papers.  He is the organizer and founder of the research team **"Project6thSense"** Their Team Aimed at stimulating the human brain started in 2009 and did many successful activities. 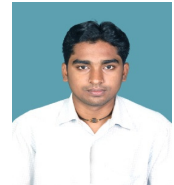

His research areas include artificial intelligence, global warming, image processing, Database management system, Data mining, humanoid, bioinformatics, Cloud computing, etc.
Website: www.indianid.6thsense.us              Reach us:  project6thsense@googlegroups.com

**Saravana Kumar S G** Born in 1988 in Kumbakonam, he did B.Sc in SASTRA University, kumbakonam, Tamil Nadu during the year (2006-09) and he going to pursing MCA in SASTRA University, kumbakonam, Tamil Nadu during the year (2009-11). He has done 9 research papers on the field NOSQL, Robotics, Green computing, Cloud Computing and two among them went to international standards one was published by IEEE. He is one of the key researcher of the research team "Project6thSense". 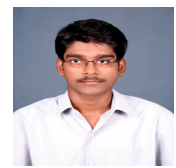